\documentclass[showpacs,twocolumn,amsfonts,amsmath,amssymb,pre,aps,dvips,superscriptaddress]{revtex4}

\usepackage{graphicx,epsfig}
\usepackage{ae}
\usepackage{bm}

\begin{document}

\title{Tunable diffusion of magnetic particles in a quasi-one-dimensional channel}

\author{D.~Lucena}
\email{diego@fisica.ufc.br}
\affiliation{Departamento de F\'isica, Universidade Federal do Cear\'a, Caixa Postal 6030, Campus do Pici, 60455-760 Fortaleza, Cear\'a, Brazil}
\affiliation{Department of Physics, University of Antwerp, Groenenborgerlaan 171, B-2020 Antwerpen, Belgium}

\author{W.~P.~Ferreira}
\email{wandemberg@fisica.ufc.br}
\affiliation{Departamento de F\'isica, Universidade Federal do Cear\'a, Caixa Postal 6030, Campus do Pici, 60455-760 Fortaleza, Cear\'a, Brazil}

\author{F.~F.~Munarin}
\affiliation{Departamento de F\'isica, Universidade Federal do Cear\'a, Caixa Postal 6030, Campus do Pici, 60455-760 Fortaleza, Cear\'a, Brazil}

\author{G.~A.~Farias}
\affiliation{Departamento de F\'isica, Universidade Federal do Cear\'a, Caixa Postal 6030, Campus do Pici, 60455-760 Fortaleza, Cear\'a, Brazil}

\author{F.~M.~Peeters}
\email{francois.peeters@ua.ac.be}
\affiliation{Department of Physics, University of Antwerp, Groenenborgerlaan 171, B-2020 Antwerpen, Belgium}
\affiliation{Departamento de F\'isica, Universidade Federal do Cear\'a, Caixa Postal 6030, Campus do Pici, 60455-760 Fortaleza, Cear\'a, Brazil}

\date{\today}


\begin{abstract}
The diffusion of a system of ferromagnetic dipoles confined in a quasi-one-dimensional parabolic trap is studied using Brownian dynamics simulations. We show that the dynamics of the system is tunable by an in-plane external homogeneous magnetic field. For a strong applied magnetic field, we find that the mobility of the system, the exponent of diffusion and the crossover time among different diffusion regimes can be tuned by the orientation of the magnetic field. For weak magnetic fields, the exponent of diffusion in the subdiffusive regime is independent of the orientation of the external field.
\end{abstract}


\maketitle

\section{Introduction}
The study of magnetic colloids is of great importance both from a theoretical and an experimental point of view. Recently there has been an increased interest in the study of the structural and the dynamical properties of magnetic confined (in particular on the meso- and nano-scale) systems due to the possibility of biomedical \cite{Lubbe,Dobson,Bergemann} and engineering applications \cite{KRaj}. Examples of these magnetic systems are ferrofluid nanofilms \cite{Rosensweig,Bacri,Rehberg} and magnetorheological (MR) fluids \cite{FlekkoyPRL,GastPRE}. For instance, the translational dynamics of a mesoscopic 3D system of permanent magnetic dipoles was studied in Ref.~\cite{KlappPRL2011}, and it was found that the system displays signatures of subdiffusive motion due to the strong suppression of orientational fluctuations of the magnetic dipoles by the presence of an homogenous external magnetic field. The formation of chains of magnetic dipoles (coagulation effect \cite{Morozov,JFaraudo,MartinJE}) is also relevant for the dynamical properties of these magnetic systems and may lead to different regimes of diffusion. Magnetic clusters of dipolar particles were recently investigated experimentally \cite{SnezhkoPRL94,SnezhkoPRL99,SnezhkoPRL102} and they may serve, e.g., as drug delivery mechanisms in biological applications. The structural properties of magnetic colloids were recently analyzed experimentally \cite{JPhysCondHempelmann} and by means of molecular dynamics simulations \cite{PREKlapp}, where novel field-induced structural transitions were observed in confined ferrofluid nanofilms.

In comparison with infinite 3D or 2D systems, confined systems exhibit a particular behavior due to the competition between the confining potential and the interparticle interaction potential. For instance, for a 2D system of repulsive particles confined in a circular parabolic potential, previous studies clearly identified the effect of the boundaries on the structural and dynamical properties of the system, as well as on the melting \cite{PRE6720031,PRE682003,PRB491994,PRE6720032,PRE712005,PRB511995}. Another interesting possibility of confined systems is realised when the 2D system is subjected to an external confining potential (e.g. parabolic) in one direction. The system is called quasi-one-dimensional (q1D). Such a q1D system of repulsive interacting particles self-organize in a chain-like structure that was recently studied experimentally \cite{GoreePRL,GoreePRE,DoyleLang,LeidererPRL}, and through analytical and numerical calculations \cite{DoylePRE,DoylePRE2,PiacentePRB,WandPRB,WandJPCM}.

Diffusion is strongly modified in confined systems, and may lead to single-file diffusion (SFD) \cite{TEHarris,Oliver,Keynes,WeiScience,SaintJeanPRE2010,TaloniPRE,EPLKwin,PereyraPRE,LucPRE}, which is directly related to the geometrical constrains imposed by an external confining potential. Furthermore, q1D systems can be used as models for the study of collective phenomena in low dimensional systems, e.g, vortex matter in type-II superconductors \cite{SoaresPRB,MiskoPRB2011}, colloidal particles \cite{MiskoPRE2009,MiskoPRE2010} and dusty plasmas. In addition, the mechanisms of ion transport in narrow channels \cite{TAllen} and DNA manipulation using magnetic particles \cite{Berensmeier,Wirtz} can be studied by modelling q1D systems.

In the present paper we investigate numerically the properties of a system of ferromagnetic dipolar particles confined in a one-dimensional parabolic trap (which models a q1D channel) coupled to a thermal bath. The orientation and strength of an in-plane external magnetic field $\textbf{B}$ are now control parameters that are able to influence the dynamics of the particles. For diluted systems, particles are arranged in a single chain structure in the center of the parabolic channel. When $\textbf{B}$ is perpendicular to the channel, the magnetic particles interact through a pure repulsive potential. For any other orientation of $\textbf{B}$, an extra attractive term is present in the particle-particle interaction potential. The latter can be dominantly attractive or repulsive, depending on the orientation of the external magnetic field. In our numerical analysis, we perform extensive Brownian dynamics (BD) simulations and calculate the mean-square displacement (MSD) $W(t)$ of the particles for different parameters which characterizes the system. For the case of normal diffusion regime (Einstein or Fickian diffusion), one has $W(t) = D_{0}t^{\alpha}$, where $D_{0}$ is the ``free particle'' diffusion coefficient, $\alpha$ is the so-called exponent of diffusion (in this case, $\alpha = 1.0$) and $t$ is time. For values of $\alpha \neq 1.0$, diffusion is said to be anomalous. For instance, in the case of SFD, $W(t) = 2Ft^{\alpha}$ (with $\alpha=0.5$) where $F$ is the single-file diffusion mobility factor. We show that the application of an in-plane homogeneous external magnetic field leads to different regimes of diffusion depending on the orientation and strength of the field.

We emphasize here that our analysis of the exponent of diffusion ($\alpha$) is restricted to the intermediate regime (ITR), which is found before the onset of the true ``long-time'' limit (i.e. $t \rightarrow \infty$) \cite{KollmannPRL}. See also discussion in Ref.~ \cite{LucPRE} and references therein. Note that in the limit $t \rightarrow \infty$, the MSD $W(t) \propto t^{0.5}$ for any pairwise interaction potential if the system fulfills the SF (single-file) condition, i.e, no particle crossings are allowed. The reason is that the clustering of particles, observed in our work due to the attractive interaction, can be considered as a system of bigger particles with lower effective particle density and smaller diffusion constant. These clusters should have the MSD $W(t) \propto t^{0.5}$ but now at a much larger time scale, which we do not consider in this work.

This paper is organized as follows. In Sec. II we introduce the model system and the numerical methods used in our study, including the approximations and limitations of our model. In Sec. III we discuss the different interaction regimes of our system. The dynamics in the strong magnetic field case is analyzed in Sec.~IV. The weak magnetic field case is discussed in Sec. V and in Sec. VI, we further investigate the strength of the magnetic field on diffusion. Finally, we present the conclusions of our work in Sec. VII.

\section{Model and Numerical Methods}\label{Model}

\subsection{Model System}
\begin{figure}[ht]
\begin{center}
\includegraphics[width=7.0cm]{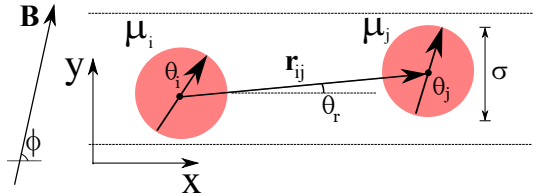}
\end{center}
\caption{(Color online) Schematic representation of the system. The particles have diameter $\sigma$ and dipole moment $\bm{\mu}_{i}$, which forms an angle $\theta_{i}$ with respect to the $x$-axis. An in-plane external magnetic field $\textbf{B}$ is applied with magnitude $B$ and $\phi$ is the angle between $\textbf{B}$ and the $x$-axis.}\label{modelFig}
\end{figure}

Our system consists of $N$ interacting dipolar ferromagnetic particles confined in a quasi-one-dimensional (q1D) channel and which is in contact with a thermal bath at absolute temperature $T$. The pair interaction potential $V_{\text{pair}}(r)$ is given by the sum of the dipole-dipole term $V_{\text{dip}}(r)$ and the short-range repulsion $V_{\text{ss}}(r)$, such as
\begin{equation}
V_{\text{pair}}(r_{ij}) = \frac{\bm{\mu}_{i} \cdot \bm{\mu}_{j}}{|\textbf{r}_{ij}|^{3}} - \frac{3(\bm{\mu}_{i} \cdot \textbf{r}_{ij})(\bm{\mu}_{j} \cdot \textbf{r}_{ij})}{|\textbf{r}_{ij}|^{5}} + 4\varepsilon \left( \frac{\sigma}{|\textbf{r}_{ij}|} \right)^{12},
\end{equation}
where $\textbf{r}_{ij}$ is the interparticle separation vector between a pair of particles $i$ and $j$, $\bm{\mu}_{i}$ is the permanent magnetic moment of particle $i$, $\sigma$ is the diameter of each particle and $\varepsilon$ is an energy parameter which characterizes the short-range repulsion between the particles and prevent them from coalescing in a single point \cite{MunarinPREdips}. We assume identical particles, i.e., $|\bm{\mu}_{i}| = |\bm{\mu}_{j}| = \mu$. The q1D channel is modeled by a parabolic confinement potential defined as $V_{\text{conf}} = m \omega^{2}y^{2}_{i}/2$, where $m$, $\omega$ and $y_{i}$ are the mass of each particle, the confinement strength (frequency) and the $y$ coordinate of the $i$th particle, respectively. We also apply an in-plane homogenous external magnetic field $\textbf{B}$, which forms an angle $\phi$ with respect to the $x$-axis. The interaction torque $\bm{\tau}_{i}$ between particles is given by $\bm{\tau}_{i} = \bm{\mu}_{i} \times \sum_{j>i}\textbf{B}^{\text{int}}_{ij}$ (see Appendix). The coupling between the magnetic moment of each particle and the external field is given by $\bm{\tau}^{B}_{i} = \bm{\mu}_{i} \times \textbf{B}$. In Fig.~\ref{modelFig}, we show a schematic representation of the system under study together with the relevant parameters.

We assume that the motion of the particles is overdamped which is typical for colloids moving in a liquid. The equations of motion for the $i$th magnetic dipolar particle are
\begin{eqnarray}\label{eqMotion2}
\gamma \dot{\textbf{r}}_{i} &=& -\sum_{j>i} [\bm{\nabla}_{i} (V_{\text{dip}} + V_{\text{ss}})] - \bm{\nabla}_{i} V_{\text{conf}} + \bm{\xi}_{i}(t),\label{eqMov1} \\
\gamma\sigma^{2} \dot{\theta_{i}} \hat{\textbf{z}} &=& \bm{\tau}_{i} + \bm{\tau}^{B}_{i} + \sigma \xi_{i}(t)\hat{\textbf{z}}\label{eqMov2},
\end{eqnarray}
where $\textbf{r}_{i} = x_{i}\hat{\textbf{x}} + y_{i}\hat{\textbf{y}}$ is the position vector of particle $i$ and $\theta_{i}$ is the angle between the vector $\bm{\mu}_{i}$ and the $x$-axis. Furthermore, $\gamma$ is the viscosity of the medium and $\bm{\xi}_{i}(t)$ is a stochastic white-noise with the properties: (i) $\langle \bm{\xi}_{i}(t) \rangle = 0$ and (ii) $\langle \xi_{im}(t)\xi_{jn}(t') \rangle = 2\gamma k_{B}T \delta_{ij} \delta_{mn} \delta(t-t')$, where $m,n$ corresponds to the components $(x,y,\theta)$, $k_{B}$ is the Boltzmann constant and $T$ is the absolute temperature of the system.

\begin{figure}[ht]
\begin{center}
\includegraphics[width=8.0cm]{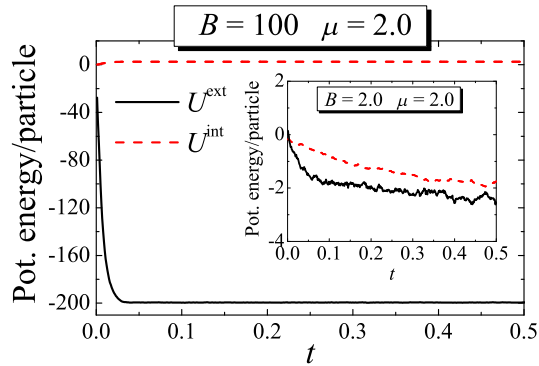}
\end{center}
\caption{(Color online) Potential energy, as defined by Eqs.~(\ref{PotEneInt})-(\ref{PotEneExt}), per particle as a function of time $t$ for $B=100$, $\mu = 2.0$. In the inset we show the same, but for $B=2$. In both cases, the number of particles in the computational unit cell was $N=300$ and all the other parameters are given in Sec.~\ref{NumMet}.}\label{figUeff}
\end{figure}

Note that the first and the second term on the rhs (right-hand side) of Eq.~(\ref{eqMov2}) are related to the potential energy of a dipole due to the magnetic field generated by all the other dipoles
\begin{equation}
U^{\text{int}} = -\bm{\mu}_{i} \cdot \sum_{j>i} \textbf{B}^{\text{int}}_{ij},\label{PotEneInt}
\end{equation}
and the potential energy of a dipole in the presence of the external magnetic field
\begin{equation}
U^{\text{ext}} = -\bm{\mu}_{i} \cdot \textbf{B},\label{PotEneExt}
\end{equation}
respectively. Therefore, for the case of a strong magnetic field (in the following we consider $B=100$ as an example), the effect of the interaction torque $\bm{\tau}_{i}$ can be neglected since the dipoles will tend to align completely to the external field, i.e., $U^{\text{int}} + U^{\text{ext}} \approx U^{\text{ext}}$ (see main panel of Fig.~\ref{figUeff}). On the other hand, if the external magnetic field is weak (for example, $B=2.0$), the interaction torque $\bm{\tau}_{i}$ can not be neglected since, for this case, we have $U^{\text{int}} \approx U^{\text{ext}}$ (see inset of Fig.~\ref{figUeff}). Nevertheless, in all our simulations we keep both terms, i.e., $\bm{\tau}_{i}$ and $\bm{\tau}^{B}_{i}$.

Finally, our model system does not take into account hydrodynamic interaction (HI) effects (particle-fluid and particle-wall interactions), which usually have only a small effect on the qualitative behavior of the diffusion properties, as recently demonstrated by Eu\'an-D\'iaz \textit{et al.} \cite{EdithPRE}. A similar approach was adopted for a dilute dipolar colloidal suspension in Refs.~\cite{RubioPRE,RubioJNNFM}, where, similar to our work, the interaction potential between particles had both a repulsive and an attractive term. The HI effects can be neglected in our case because we are in the dilute regime, i.e., the low density case. Note that the particles are almost completely uniformly distributed along the $x$-direction, i.e., the system forms a single-chain configuration. Furthermore, HI effects should play an important role in diffusion (and in general, in dynamical properties) for the case of highly concentrated colloidal suspensions \cite{GNagele}, a situation that is not considered in our work.


\subsection{Numerical Methods}\label{NumMet}
Before we integrate numerically Eqs.~(\ref{eqMov1}) and (\ref{eqMov2}), we introduce the unit of time as $t_{0} = \sigma^{2}\gamma/\varepsilon$, where $\varepsilon = k_{B}T_{0}$ is the unit of energy ($T_{0}$ is the unit of temperature) and $\sigma$ is the unit of length. Moreover, $B_{0} = \sqrt{\varepsilon/\sigma^{3}}$ is the unit of magnetic field and $\mu_{0} = \sqrt{\varepsilon\sigma^{3}}$ is the unit of magnetic moment, $\omega_{0} = (t_{0})^{-1}$ and the dimensionless parameter $\omega^{*} = m(\omega\sigma)^{2}/2\varepsilon$ controls the strength of the parabolic confinement potential in the $y$-direction. These scaling turn all quantities into dimensionless (asterisk) form.

Integrating the dimensionless overdamped equations of motion, we obtain the following Ermak-type algorithm \cite{Ermak} for updating the position ($\textbf{r}^{*}_{i}$) and angle ($\theta^{*}_{i}$) of particle $i$ during the simulation time step $\Delta t^{*}$:

\begin{eqnarray}
\textbf{r}^{*}_{i}(\Delta t^{*}) &=& \textbf{r}^{*}_{i}(0) + \Delta t^{*} \textbf{f}^{*}_{ij} + \Delta t^{*} (\omega^{*})^{2} \textbf{g}^{*}_{i} + \sqrt{2T^{*}\Delta t^{*}} \bm{\xi}^{*}_{i}, \nonumber \\
\theta^{*}_{i}(\Delta t^{*}) &=& \theta^{*}_{i}(0) + \Delta t^{*}\tau^{*}_{i} + \Delta t^{*} \tau^{*B}_{i} + \sqrt{2T^{*}\Delta t^{*}} \xi^{*}_{i},
\end{eqnarray}
where $\textbf{f}^{*}_{ij} = - \sum_{j} \bm{\nabla}^{*}_{i}[V^{*}_{\text{dip}} + V^{*}_{\text{ss}}]$, $\textbf{g}^{*}_{i} = - \bm{\nabla}^{*}_{i}[(y^{*}_{i})^{2}]$, $\tau^{*}_{i} = |\bm{\mu}^{*}_{i} \times \sum_{j>i}\textbf{B}^{*\text{int}}_{ij}|$ (see Appendix) and $\tau^{*B}_{i} = |\bm{\mu}^{*}_{i} \times \textbf{B}^{*}|$. Furthermore, $V^{*}_{\text{dip}}$ and $V^{*}_{\text{ss}}$ are given by
\begin{eqnarray}
V^{*}_{\text{dip}} &=& \frac{\bm{\mu}^{*}_{i} \cdot \bm{\mu}^{*}_{j}}{|\textbf{r}^{*}_{ij}|^{3}} - \frac{3(\bm{\mu}^{*}_{i} \cdot \textbf{r}^{*}_{ij}) (\bm{\mu}^{*}_{j} \cdot \textbf{r}^{*}_{ij})}{|\textbf{r}^{*}_{ij}|^{5}}, \\
V^{*}_{\text{ss}} &=& 4/|\textbf{r}^{*}_{ij}|^{12}.
\end{eqnarray}


From this point onward we will abandon the asterisk notation and all physical quantities are dimensionless, unless stated otherwise. In our simulations, we use the following parameters: $\Delta t = 1.0 \times 10^{-6}$, $\omega = 10.0$, $\mu = 2.0$ and $T=1.0$. Note that $B$ and $T$ can be related by the dimensionless parameter $\eta = |U^{\text{eff}}|/k_{B}T$, which is defined as the ratio between the coupling energy of a dipole particle with the effective magnetic field ($U^{\text{eff}}=U^{\text{int}}+U^{\text{ext}}$) and the thermal energy ($k_{B}T$). We also use a simulation box of length (in the $x$ direction) $L_{x} = 375.0$, and linear density $\rho = N/L_{x} = 0.8$. We choose this value of $L_{x}$ in order to cutoff the interaction potential for distances larger than $r = r_{c} = L_{x}/2 \approx 187.0$, at which the interaction energy between a pair of particles is approximately $V_{\text{dip}}(r)|_{r_{c}} \approx 1.0 \times 10^{-6}$. In the $x$ direction, we apply periodic boundary conditions and in the transverse direction, the system is confined by the parabolic trap, which is controlled by the parameter $\omega$. Note that in this work we set a value of $\omega$ which is large enough to prevent particles from bypassing each other, as we demonstrated in a previous study \cite{LucPRE}. This forces the system into a strict 1D chain of particles in the $x$ direction. The initial configuration of the particles is chosen randomly and the system is equilibrated during $(1.0-5.0) \times 10^{6}$ simulation time steps. Other parameters which characterize the system are the magnitude of the external magnetic field ($B$) and the angle $\phi$ between $\textbf{B}$ and the $x$-axis. Furthermore, the stochastic white noise $\bm{\xi}_{i}(t)$ is simulated using the Box-M\"{u}ller transformation technique \cite{BoxMuller} and in all the results presented in this work, the error bars in the plots are smaller than the symbol size.


\section{Interaction potential between two dipoles}\label{1dchain}
\begin{figure}[ht]
\begin{center}
\includegraphics[width=8.0cm]{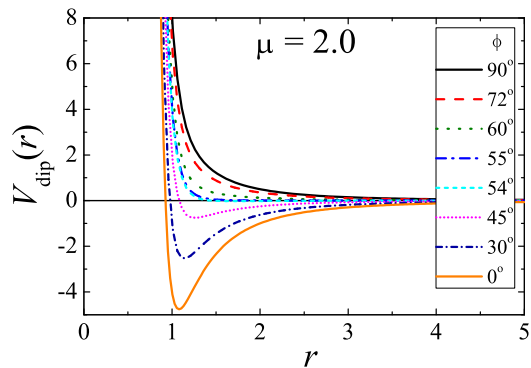}
\end{center}
\caption{(Color online) Dipole-dipole interaction potential $V_{\text{dip}}(r)$ [Eq.~(\ref{vdipEq})] as a function of the distance $r$ between two dipoles and for different values of $\phi$.}\label{vdip}
\end{figure}

Before we study the complete system (i.e., the model described in Sec.~\ref{Model}), let us first analyze the behavior of the dipole-dipole interaction potential $V_{\text{dip}}(r)$ between two particles as a function of $\phi$ (cf. Fig.~\ref{modelFig}), assuming that both dipoles are perfectly oriented in the direction of the external field. In this case, the interaction potential may be written as
\begin{equation}\label{vdipEq}
V_{\text{dip}}(|\textbf{r}|) = \frac{|\bm{\mu}|^{2}}{|\textbf{r}|^{3}} \left[1 - 3\cos^{2}(\phi - \theta_{r}) \right] + 4|\textbf{r}|^{-12},
\end{equation}
where $\theta_{r}$ (cf. Fig.~\ref{modelFig}) is the angle formed between the vector $\textbf{r}$ and the $x$-axis. We assume the simplest case, where $\theta_{r} = 0^{o}$, which means that particles are forming a perfect one-dimensional chain along the $x$ direction. The dependence of $V_{\text{dip}}$ [Eq.~(\ref{vdipEq})] on the distance $r$ between two particles is presented in Fig.~\ref{vdip} for different values of $\phi$. We found that for $\phi \gtrapprox 54^{o}$, the interaction potential is dominantly repulsive. On the other hand, for $\phi \lessapprox 54^{o}$, the interaction potential has a Lennard-Jones form (e.g, $\phi=0^{o}$ in Fig.~\ref{vdip}). For small values of $r$, the repulsive term $4|\textbf{r}|^{-12}$ is dominant. For intermediate values of $r$ ($1.0 < r <1.5$), the particle can be trapped in the potential well due to the presence of the attractive part in the interaction potential. For larger distances ($r \rightarrow \infty$), the interaction vanishes.

\section{Influence of a strong external magnetic field on diffusion}\label{extB100}
The influence of a strong homogeneous external magnetic field on the diffusive properties of the model system described in Sec.~\ref{Model} will now be investigated. The external field $\textbf{B}$ with magnitude $B = 100$ (which is a typical strong field value used in experiments, see e.g. \cite{KlokPRL}) forms an angle $\phi$ with respect to the $x$-axis (cf. Fig.~\ref{modelFig}). Note that since we set $T=1.0$, the parameter $\eta \approx 200 \gg 1$, which means thermal fluctuations are weak. We now investigate how diffusion depends on $\phi$.

We will study the diffusive properties of the system through the analysis of the mean-square displacement $W(t)$ along the $x$ direction, defined as
\begin{eqnarray}\label{MSDsystem}
W(t) = \left\langle \frac{1}{N} \sum_{i=1}^{N} [x_{i}(\tau + \delta t) - x_{i}(\tau)]^{2} \right\rangle_{\tau} ,
\end{eqnarray}
where $N$ is the number of particles (we use a typical value of $N$=300--900 particles), $\tau$ is an arbitrary time origin \cite{SaintJeanPRE}, $\delta t$ is the time interval between measurements and $\langle \cdot \rangle_{\tau}$ is an average over different time origins during the simulation \cite{FrenkelUNDER}.

\subsection{Region (I): $55^{o} \lesssim \phi \leq 90^{o}$}\label{RegI}
First, we consider the external magnetic field perpendicular to the parabolic confinement channel, i.e. $\phi = 90^{o}$. In this case, the interaction is purely repulsive [i.e. $V_{\text{dip}}(r) \propto (1/r)^{3}$] and the mean-square displacement $W(t)$ [Fig.~\ref{fig:msdB100}(a)] of the system exhibits a subdiffusive regime [single-file diffusion (SFD)], with $W(t) = 2F_{a}t^{0.5}$ for time scales larger than the short-time normal diffusion regime (STND), which is characterized by $W(t) = D_{0}t$ \cite{HerreraSuperParaPRE}. The crossover time $t_{c}$ between these two distinct regimes of diffusion can be estimated \cite{SaintJeanPRE} as the time where the curves $D_{0}t$ and $2F_{a}t^{0.5}$ intersect:
\begin{equation}\label{tcross}
D_{0}t_{c} \approx 2F_{a}(t_{c})^{0.5} \Rightarrow t_{c} \approx \left(\frac{2F_{a}}{D_{0}}\right)^{2}.
\end{equation}

The mean-square displacement for $\phi = 90^{o}$ and $\phi = 70^{o}$ are presented in Figs.~\ref{fig:msdB100}(a)-(b), respectively. We found that for $55^{o} \lesssim \phi \leq 90^{o}$, i.e., when the dipole-dipole interaction is purely repulsive (cf. Fig.~\ref{vdip}), $W(t)$ has the following behavior
\begin{equation}
W(t) = \left\{
\begin{array}{ll} D_{0}t & \mbox{for } t < t_{c}  \\
2F_{a}t^{0.5} & \mbox{for } t > t_{c},
\end{array}
\right.
\end{equation}
where a straightforward calculation using Eq.~(\ref{tcross}) gives $t_{c} \approx 7.58 \times 10^{-3}$ ($F_{a} \approx 4.79 \times 10^{-5}$ and $D_{0} \approx 0.110 \times 10^{-5}$). In this region (I), the crossover time $t_{c}$ and the SFD mobility $F_{a}$ are independent of the value of $\phi$.

\subsection{Region (II): $0^{o} \leq \phi \lesssim 55^{o}$}\label{RegII}
\begin{figure}[!h]
\begin{center}
\includegraphics[width=8.0cm]{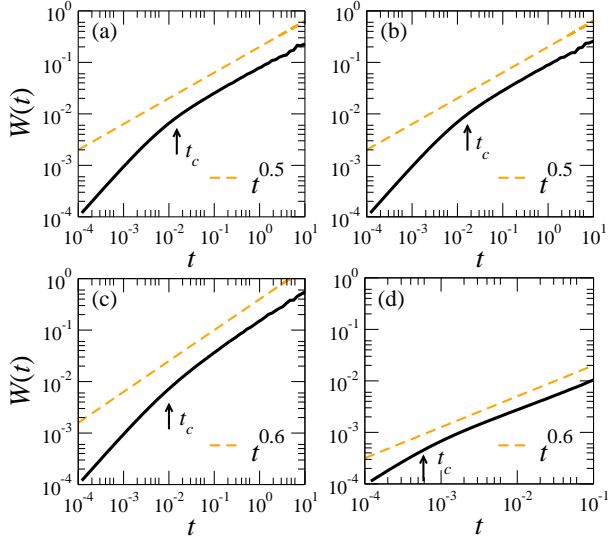}
\caption{(Color online) Log-log plot of the mean-square displacement (solid black curves) $W(t)$ as a function of the time $t$ for $B=100$ and (a) $\phi=90^{o}$, (b) $\phi = 70^{o}$, (c) $\phi = 50^{o}$ and (d) $\phi = 0^{o}$. The dashed orange lines are a guide to the eye and the crossover time $t_{c}$ for each case is indicated by the vertical arrow.}\label{fig:msdB100}
\end{center}
\end{figure}

For the case of $0^{o} \leq \phi \lesssim 55^{o}$, the attractive term present in the dipole-dipole interaction potential becomes more relevant with decreasing $\phi$. As a consequence, we expect that the diffusion of the dipoles to be affected by the orientation of $\textbf{B}$. We found that for this region (II), the system exhibits the STND followed by a subdiffusive regime, with $W(t) = 2F_{b}(\phi)t^{0.6}$, where now $t_{c}$ and $F_{b}$ depends on the angle $\phi$ and
\begin{equation}
W(t) = \left\{
\begin{array}{ll} D_{0}t & \mbox{for } t < t_{c}(\phi)  \\
2F_{b}(\phi)t^{0.6} & \mbox{for } t > t_{c}(\phi),
\end{array}
\right.
\end{equation}
with $t_{c}(\phi) \approx (2F_{b}(\phi)/D_{0})^{2.5}$. The mean-square displacement for $\phi = 50^{o}$ and $\phi = 0^{o}$ is presented in Figs.~\ref{fig:msdB100}(c)-(d), respectively.

\begin{figure}[!h]
\begin{center}
\includegraphics[width=8.0cm]{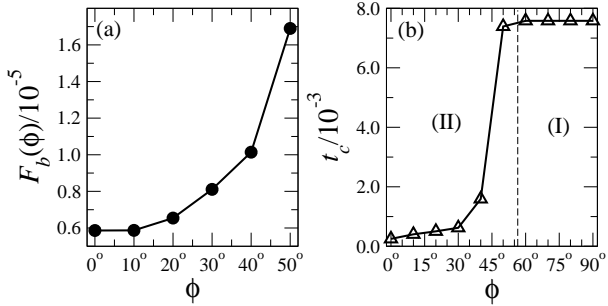}
\caption{(a) Mobility $F_{b}$ in region (II) as a function of $\phi$ and (b) crossover time $t_{c}$ between the STND regime and the subdiffusive regime as a function of $\phi$. The solid lines are a guide to the eye. The dashed vertical line in (b) divides regions with (II) and without (I) an attractive part in the inter-particle interaction potential.}\label{fig:FbtcB100}
\end{center}
\end{figure}

In Figs.~\ref{fig:FbtcB100}(a)-(b) we show the mobility $F_{b}(\phi)$ in region (II) and the crossover time $t_{c}$ as a function of $\phi$, respectively. Note that both $F_{b}$ and $t_{c}$ decreases with decreasing $\phi$ in region (II). On the other hand, as stated above, the crossover time $t_{c}$ is constant in region (I).

The decrease of $t_{c}$ and $F_{b}$, in region (II), with decreasing $\phi$ can be explained by the decrease of the minimum interparticle distance between neighbor particles [cf. Fig.~\ref{alphadistminB100}(a)]. When the interaction potential is dominated by the repulsive part of the potential [i.e., region (I)], the particles are distributed homogenously along the unconfined direction [Fig.~\ref{snapB100}(a)], i.e., the minimum interparticle distance between neighbors is approximately constant. In region (II), the attractive term in the interaction potential becomes more relevant, and the system starts to form clusters of chains. Therefore, the particles are no longer homogenously distributed along the channel. The minimum interparticle distance decreases with decreasing $\phi$ and the crossover time $t_{c}$ is smaller than in region (I) because particles ``feel'' the interaction with neighboring particles much faster. Also, since the particles can be trapped inside the clusters of chains, the mobility ($F_{b}$) is reduced with decreasing $\phi$.

\subsection{Exponent of diffusion ($\alpha$) in the intermediate (ITR) subdiffusive regime}\label{ExpAlpha}

\begin{figure}[t]
\begin{center}
\includegraphics[width=6.0cm]{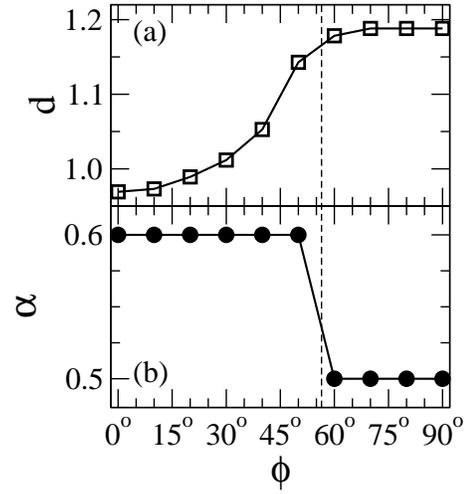}
\end{center}
\caption{(a) Minimum interparticle distance $d$ between neighboring particles for $B=100$ and $T=1.0$ as a function of the orientation $\phi$ of the external field. (b) Exponent of diffusion ($\alpha$) as a function of the orientation $\phi$ of the external magnetic field. Note that $d$ decreases with decreasing $\phi$ in the region $0^{o} < \phi \lesssim 55^{o}$, which is the same region where we found the increase of the diffusion mechanism [cf. panel (b)]. The solid lines are a guide to the eye.}\label{alphadistminB100}
\end{figure}

\begin{figure}[!ht]
\begin{center}
\includegraphics[width=7.0cm]{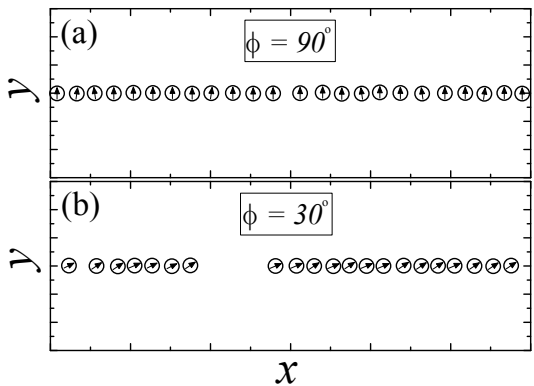}
\end{center}
\caption{Typical snapshots of the system after $10^{6}$ simulation time steps for (a) $\phi = 90^{o}$ and (b) $\phi = 30^{o}$. Other parameters are $B = 100$ and $T = 1.0$.}\label{snapB100}
\end{figure}

In the previous section we showed that the MSD [$W(t)$] exhibits two different regimes of subdiffusion depending on the region [(I) or (II)]: the exponent of diffusion ($\alpha$) in the subdiffusive regime changes from $\alpha=0.5$ to $\alpha=0.6$ as the angle $\phi$ is decreased from $\phi=90^{o}$ to $\phi=0^{o}$. The exponent $\alpha$ is calculated by fitting the MSD of our simulation data in the region of interest (for instance, the ITR regime) according to the relation $W(t) \propto t^{\alpha}$. The increase in the diffusion mechanism can be seen in Fig.~\ref{alphadistminB100}(b), where $\alpha$ is presented as a function of the orientation $\phi$. Note that $\alpha$ increases with decreasing $\phi$, which can be understood in terms of the dipole-dipole interaction dependence on $\phi$. For $\phi \gtrsim 55^{o}$, the interaction potential is mainly repulsive and therefore it leads the system into a subdiffusive behavior, where $\alpha = 0.5$. The scaling $W(t) \propto t^{0.5}$ has been observed experimentally in repulsive interacting particles \cite{WeiScience}, and was also found from simulations \cite{MiskoPRE2010,SaintJeanPRE} and through analytical \cite{Lizana,PMCentres} calculations. In this case, the minimum interparticle distance is approximately equal to $d \approx (\rho)^{-1} \approx 1.2$. On the other hand, for $\phi \lesssim 55^{o}$, the interaction potential exhibits a competition between a repulsive and an attractive term (cf. Fig.~\ref{vdip}). The attractive part of the potential forces the formation of clusters of chains [Fig.~\ref{snapB100}(b)], resulting in empty spaces along the unconfined direction. This is illustrated in Fig.~\ref{alphadistminB100}(a), where the minimum distance between particles $d$ is shown as a function of $\phi$. Note that $d$ decreases with decreasing $\phi$. Since the system has a fixed density $\rho$, the empty spaces between the clusters of chains results in an increase of diffusion, which subsequently gives an exponent of diffusion $\alpha$ that is slightly larger than $0.5$.

\begin{figure}[!ht]
\begin{center}
\includegraphics[width=8.0cm]{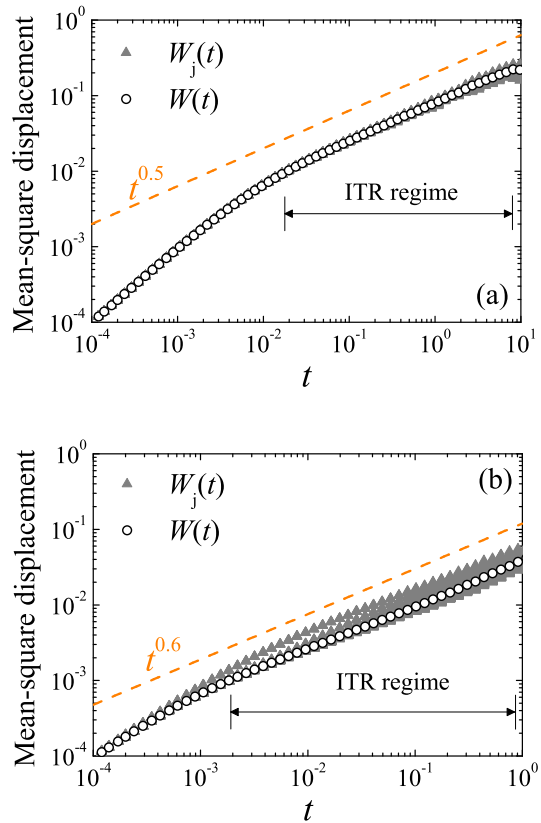}
\end{center}
\caption{(Color online) Mean-square displacement of the system [open black circles, $W(t)$] and mean-square displacement of individual particles [gray triangles, $W_{j}(t)$] as a function of the time $t$ for two different values of $\phi=$ (a) $90^{o}$ and (b) $0^{o}$. The dashed orange lines are a guide to the eye. Other parameters are $B=100$ and $T=1.0$.}\label{indMSDB100}
\end{figure}

In order to better understand the increase of the exponent of diffusion $\alpha$, we calculate the mean-square displacement of each $j$th particle [$W_{j}(t)$] using an expression similar to Eq. (\ref{MSDsystem}):
\begin{eqnarray}\label{MSDpart}
W_{j}(t) = \langle [x_{j}(\tau + \delta t) - x_{j}(\tau)]^{2} \rangle_{\tau} ,
\end{eqnarray}
where $j=1,...,N$ represents each individual particle and $\langle \cdot \rangle_{\tau}$ is an average over different time origins during the simulation. In Figs.~\ref{indMSDB100}(a)-(b) we show $W(t)$ (open black circles) and $W_{j}(t)$ (gray triangles) for $\phi=90^{o}$ and $\phi=0^{o}$. Note that for the case $\phi=90^{o}$, $W_{j}(t)$ deviates very little from the mean-square displacement of the system $W(t)$. In this case the particles in the system are distributed homogenously along the unconfined direction. Therefore, the diffusion of a tagged particle should be the same as the diffusion of the whole system. On the other hand, for the case of $\phi=0^{o}$, $W_{j}(t)$ deviates [much more] from $W(t)$ [than in the case $\phi=90^{o}$]. This is caused by the asymmetry along the unconfined direction. In this case, it is possible that a tagged particle can diffuse differently than the whole system because of the formation of clusters of chains [cf. Fig.~\ref{snapB100}(b)]. For instance, particles which are located at the borders of the cluster of chains diffuse faster than particles which are inside the cluster. This is the reason for an exponent $\alpha$ that is slightly larger than $0.5$ in the case where the interaction potential has both repulsive and attractive terms [Region (II), see Sec.~\ref{RegII}].

\section{Weak magnetic fields}
\begin{figure}[!h]
\begin{center}
\includegraphics[width=8.0cm]{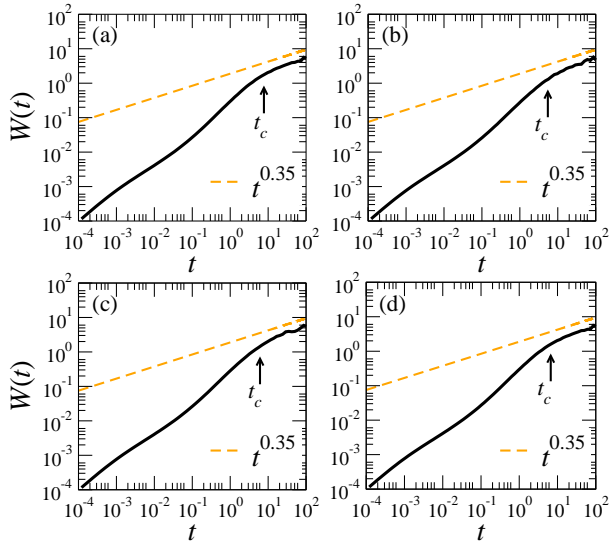}
\caption{(Color online) Log-log plot of the mean-square displacement (solid black curves) $W(t)$ as a function of the time $t$ for $B=0.1$ and (a) $\phi=90^{o}$, (b) $\phi = 70^{o}$, (c) $\phi = 50^{o}$ and (d) $\phi = 0^{o}$. The dashed orange lines are a guide to the eye and the crossover time $t_{c}$ for each case is indicated by the vertical arrow.}\label{fig:msdB01}
\end{center}
\end{figure}

In the previous section, we showed that the diffusion mechanism of the system is affected by the orientation of the strong external magnetic field. Now, we turn to the question of how the magnitude of $\textbf{B}$ influences the diffusive properties of the system. To this end, we perform similar simulations using the same parameters of the previous section, but with a weaker magnetic field $B = 0.1$. Note that since we set $T=1.0$, the parameter $\eta \approx 0.2 \ll 1$, which means thermal fluctuations are strong. The mean-square displacement (in log-log scale) as a function of the time is presented in Fig.~\ref{fig:msdB01} for different values of $\phi$.

\begin{figure}[!ht]
\begin{center}
\includegraphics[width=8.0cm]{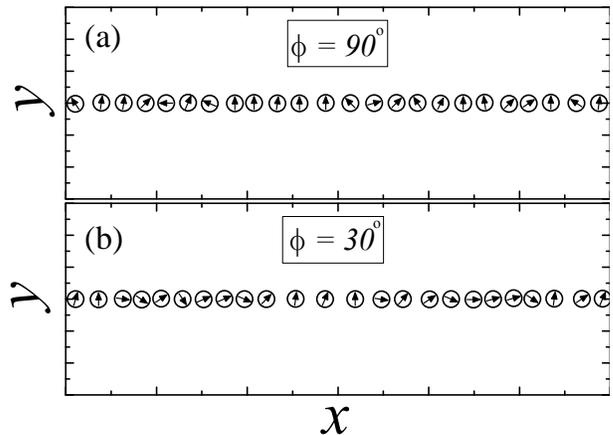}
\end{center}
\caption{Typical snapshots of the system after $10^{6}$ simulation time steps for (a) $\phi = 90^{o}$ and (b) $\phi = 30^{o}$. Other parameters are $B = 0.1$ and $T = 1.0$.}\label{snapB1}
\end{figure}

There are two important observations regarding the results for $B = 0.1$: (i) note that the ITR regime for this case is shifted to larger time intervals as compared to the previous case (see Fig.~\ref{fig:msdB100}), which is a consequence of the weaker coupling of the dipoles with the external magnetic field, leading the system to larger relaxation (crossover) times. Here, the ITR regime can be identified in the time interval $10^{1} \lesssim t < 10^{2}$; (ii) since the external magnetic field is small (compared to the case of the previous section, $B = 100$), the coupling between the magnetic dipoles and the external field is weaker which results in an approximately $\phi$-independent regime of diffusion [cf. Fig.~\ref{fig:msdB01}]. This means that the exponent of diffusion $\alpha$ in the ITR regime is a constant ($\alpha=0.35$) which is independent of the orientation of the external magnetic field. We will further discuss this particular value of $\alpha$ in the following section. Note that, as opposed to the case of strong magnetic field, the clustering of particles in a chain-like configuration along the unconfined direction is less pronounced, as illustrated in Fig.~\ref{snapB1}. Note that the orientation of the dipoles of the ferromagnetic particles is almost random.

\section{Influence of the strength of the magnetic field}
\begin{figure}[!h]
\begin{center}
\includegraphics[width=8.0cm]{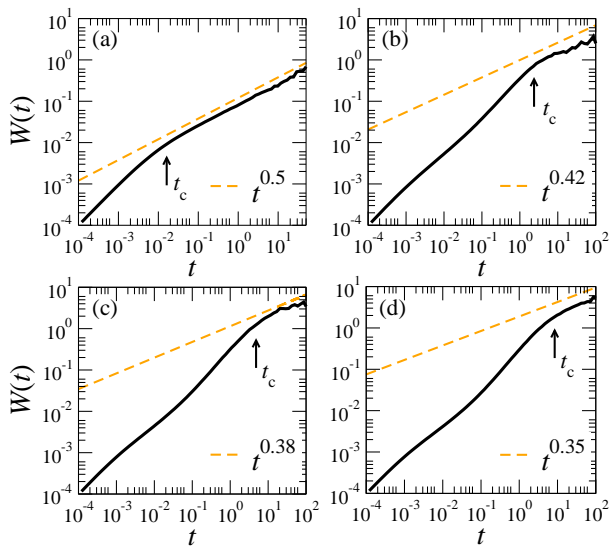}
\caption{(Color online) Log-log plot of the mean-square displacement (solid black curves) $W(t)$ as a function of the time $t$ for $\phi=90^{o}$ and (a) $B=10$, (b) $B=2$, (c) $B=1$ and (d) $B=0.1$. The dashed orange lines are a guide to the eye.}\label{MSDf90ChangeB}
\end{center}
\end{figure}

In this section, we further investigate how the strength $B$ of the external magnetic field influences the diffusion of the system. We analyze the case for $\phi=90^{o}$, where the SFD is found in the ITR regime for $B=100$ [see Fig.~ \ref{alphadistminB100}(b)]. From the calculations of the MSD using Eq.~(\ref{MSDsystem}) for different values of $B$, we found that for $B \gtrsim 10$, the SFD regime is always present in the ITR regime, i.e., $W(t) \propto t^{0.5}$. Therefore, we only investigate the region $0.1 \leq B \leq 10.0$, and the results are plotted in Figs.~\ref{MSDf90ChangeB}(a)-(d). For $B=10$ [Fig.~\ref{MSDf90ChangeB}(a)], as stated above, the SFD regime is present in the ITR regime, which means $\alpha=0.5$.

\begin{figure}[!h]
\begin{center}
\includegraphics[width=5.0cm]{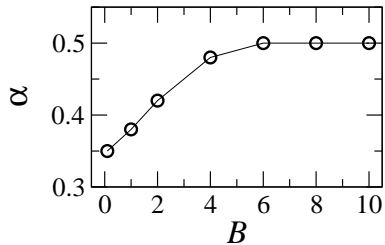}
\caption{Exponent of diffusion $\alpha$ (in the ITR regime) as a function of the strength $B$ of the external magnetic field for $\phi=90^{o}$. The solid line is a guide to the eye.}\label{alphaChangeB}
\end{center}
\end{figure}

We found that by decreasing the value of $B$, the exponent of diffusion ($\alpha$) decreases from $\alpha=0.5$ to $\alpha=0.35$, as shown in Fig.~\ref{alphaChangeB}. The reason for this behavior is explained by the following: as the magnetic field is decreased, the coupling of it with the dipoles also decreases, leading to an increase in the rotational movement of the dipoles. Therefore, the energy of a dipole is distributed between translational and rotational motion. Recall that for large values of $B$ ($=100$), the dipoles were almost completely aligned with the field. The increase in the rotation of the dipoles thus leads to a slowing down of the translational diffusion, i.e., $\alpha$ decreases with decreasing $B$.

\begin{figure}[!h]
\begin{center}
\includegraphics[width=8.0cm]{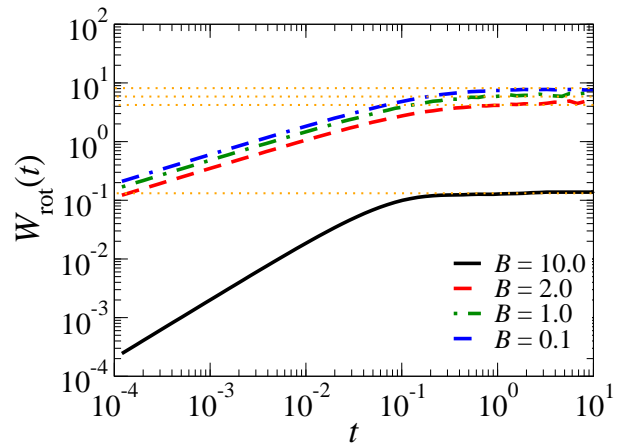}
\caption{(Color online) Log-log plot of the mean-square angular displacement $W_{\text{rot}}(t)$ as a function of the time $t$ for $\phi=90^{o}$ and $B=10$, $B=2$, $B=1$ and $B=0.1$. The dotted orange horizontal lines correpond to the saturation values of $W_{\text{rot}}(t)$.}\label{MSDRotf90ChangeB}
\end{center}
\end{figure}

In order to strengthen this conclusion, we calculate the mean-square angular displacement (MSAD) $W_{\text{rot}}(t)$, which is defined similary to Eq.~(\ref{MSDsystem}):
\begin{equation}\label{MSDsystemRot}
W_{\text{rot}}(t) = \left\langle \frac{1}{N} \sum_{i=1}^{N} [\theta_{i}(\tau + \delta t) - \theta_{i}(\tau)]^{2} \right\rangle_{\tau} ,
\end{equation}
where $\theta_{i}$ is the angular coordinate of the $i$th particle (cf. Fig~\ref{modelFig}). The results of calculations of the MSAD are shown in Fig.~\ref{MSDRotf90ChangeB} for different values of $B$. Note that for all values of $B$, $W_{\text{rot}}(t)$ saturates after the initial motion. Furthermore, the MSAD curves increases with decreasing $B$, which indicates that the rotational motion of the dipoles increases with decreasing strength of the external magnetic field.

\section{Conclusions}
We studied a system of interacting ferromagnetic dipoles, confined in a q1D channel, that are subjected to a homogeneous external magnetic field. The analysis of the mean-square displacement $W(t)$ indicates that the diffusive properties of the system depends on the orientation and on the strength of the external field. For the case of strong magnetic fields (we considered $B = 100$ as an example), we found that the exponent of diffusion $\alpha$ increases with decreasing orientation $\phi$ [cf. Fig.~\ref{alphadistminB100}(b)] of the external field, i.e., directing the magnetic field towards the direction parallel to the channel. This increase of diffusion was explained by the dependence of the dipole-dipole interaction potential on $\phi$. For $\phi \gtrsim 55^{o}$, the interaction is dominantly repulsive, leading the system into subdiffusive motion in the ITR regime. On the other hand, for $\phi \lesssim 55^{o}$, the interaction potential has a Lennard-Jones form, which creates a competition between the repulsive and the attractive term of the dipole-dipole potential. The attractive part of the potential leads the system into clusters of chains [Fig.~\ref{snapB100}(b)]. The empty spaces in the system allow for an increase in diffusion.

For small values of the magnetic field (e.g. $B = 0.1$), the coupling between the magnetic dipoles and $\textbf{B}$ is weak and the dynamic behavior of the system becomes almost independent of the orientation of $\textbf{B}$. This results in an exponent $\alpha$, in the subdiffusive regime, that is a constant ($\alpha=0.35$) for all values of the orientation of the external magnetic field. The fact that for weak magnetic fields the exponent of diffusion is smaller than $0.5$ (i.e. the slowing down of translational diffusion) was explained by the weak coupling of the dipoles with the external field, leading to an increase in the rotational motion of the dipoles. Note that the value of $\alpha=0.35$ only holds for the ITR regime, as discussed in the Introduction. In both cases (i.e. strong and weak external magnetic fields), the system is still in the single-file, diluted regime.

Our results show that the diffusion mechanism in this system can be controlled by tuning the orientation and the strength of the external magnetic field. This will allow one to control the dynamics of magnetic particles in narrow channels by simply tuning the parameters which regulate the external magnetic field.


\section*{ACKNOWLEDGMENTS}
This work was supported by CNPq, CAPES, FUNCAP (Pronex grant), the Flemish Science Foundation (FWO-Vl), the bilateral program between Flanders and Brazil, the collaborative program CNPq - FWO-Vl, and the Brazilian program Science Without Borders (CsF). Discussions with V.~R.~Misko are gratefully acknowledged.

\section*{APPENDIX}
In this section, we calculate the first ($\bm{\tau}_{i}$) and the second term ($\bm{\tau}^{B}_{i}$) present in the rhs (right-hand side) of the equation of motion (\ref{eqMov2}) in cartesian coordinates. The interaction torque $\bm{\tau}_{i}$ is given by the relation:
\begin{equation}
\bm{\tau}_{i} = \bm{\mu}_{i} \times \sum_{j>i}\textbf{B}^{\text{int}}_{ij},\label{eqTorque}
\end{equation}
where $\bm{\mu}_{i}$ is the magnetic moment of $i$th particle and $\sum_{j>i}\textbf{B}^{\text{int}}_{ij}$ is the magnetic field generated by all $j$ particles on the $i$th particle. Following Refs. \cite{JacksonClassED,KlappPRE2012}, we write:
\begin{equation}
\textbf{B}^{\text{int}}_{ij} \simeq \frac{3\hat{\textbf{n}}(\hat{\textbf{n}} \cdot \bm{\mu}_{j}) - \bm{\mu}_{j}}{|\textbf{r}_{ij}|^{3}},\label{MAGField}
\end{equation}
where $\hat{\textbf{n}}=\textbf{r}_{ij}/|\textbf{r}_{ij}|$. Since the system is (in practice) two-dimensional (2D), we may write
\begin{eqnarray}
\textbf{B}^{\text{int}}_{ij} &=& B^{x}_{ij} \hat{\textbf{x}} + B^{y}_{ij} \hat{\textbf{y}},\label{MAGFieldCart} \\
\textbf{r}_{ij} &=& \Delta x_{ij} \hat{\textbf{x}} + \Delta y_{ij} \hat{\textbf{y}},\label{vecRij} \\
\bm{\mu}_{j} &=& \mu \cos\theta_{j} \hat{\textbf{x}} + \mu \sin\theta_{j} \hat{\textbf{y}},\label{vecMuj}
\end{eqnarray}
in cartesian coordinates. Therefore, directly calculation of Eq. (\ref{eqTorque}) using Eqs. (\ref{MAGField})--(\ref{vecMuj}) gives
\begin{equation}
\bm{\tau}_{i} = \hat{\textbf{z}}\left[\mu \cos\theta_{i}\sum_{j>i}B^{y}_{ij} - \mu \sin\theta_{i}\sum_{j>i}B^{x}_{ij}\right],\label{TorqueCart}
\end{equation}
where the terms $B^{x}_{ij}$ and $B^{y}_{ij}$ are given by:
\begin{eqnarray}
B^{x}_{ij} &=& \frac{3[\Delta x^{2}_{ij}\mu\cos\theta_{j} + \Delta x_{ij}\Delta y_{ij}\mu\sin\theta_{j}] - \mu\cos\theta_{j}|\textbf{r}_{ij}|^{2}}{|\textbf{r}_{ij}|^{5}},\\
B^{y}_{ij} &=& \frac{3[\Delta x_{ij}\Delta y_{ij}\mu\cos\theta_{j} + \Delta y^{2}_{ij}\mu\sin\theta_{j}] - \mu\sin\theta_{j}|\textbf{r}_{ij}|^{2}}{|\textbf{r}_{ij}|^{5}}.
\end{eqnarray}
Similary, we can calculate the torque $\bm{\tau}^{B}_{i}$ due to the external magnetic field $\textbf{B} = B_{x}\hat{\textbf{x}} + B_{y}\hat{\textbf{y}}$ on the $i$th particle as:
\begin{equation}
\bm{\tau}^{B}_{i} = \bm{\mu}_{i} \times \textbf{B} = \hat{\textbf{z}}[\mu\cos\theta_{i}B_{y} - \mu\sin\theta_{i}B_{x}].\label{TorqueExtCart}
\end{equation}
Note that since the problem is 2D, the torques $\bm{\tau}_{i}$ and $\bm{\tau}^{B}_{i}$ [Eqs. (\ref{TorqueCart}) and (\ref{TorqueExtCart}), respectively] are in the $z$-direction, i.e., perpendicular to the $xy$-plane.

\end{document}